\documentclass{epl}

\title{Drop dynamics on chemically patterned surfaces}

\author{H. Kusumaatmaja\inst{1}\thanks{E-mail: \email{halim@thphys.ox.ac.uk}} \and J. L\'{e}opold\`{e}s\inst{2}\thanks{Present address: Laboratoire de Physico-Chimie des Polym\`{e}res, Universit\'{e} de Mons-Hainaut, 20, Place du Parc - B-7000 Mons, Belgium} \and A. Dupuis\inst{1}\thanks{Present address: Institute of computational science, ETH Zurich, 8092 Zurich, Switzerland} \and J. M. Yeomans\inst{1}}
\shortauthor{H. Kusumaatmaja \etal } 

\institute{
  \inst{1} The Rudolf Peierls Centre for Theoretical Physics, Oxford University, 1 Keble Road, Oxford OX1 3NP, UK \\
  \inst{2} Department of Materials, Oxford University, Parks Road, Oxford OX1 3PH, UK \\
  }

\date{\today}

\begin{document}

\maketitle

\begin{abstract}
We compare numerical and experimental results exploring the behaviour of liquid drops moving across a surface patterned with hydrophobic and hydrophilic stripes. A lattice Boltzmann algorithm is used to solve the hydrodynamic equations of motion of the drops allowing us to investigate their behaviour as the stripe widths and the wettability contrast are altered. We explain how the motion of the drop is determined by the interplay between the driving force and the variation in surface force as the drop moves between regions of different contact angle and we find that the shape of the drops can undergo large periodic deviations from spherical. When compared, the numerical results agree well with experiments on micron--scale drops moving across substrates patterned by microcontact printing.
\end{abstract}

\section{Introduction}

The question of how liquid drops wet and move across a solid surface has long caught the interests of academic and industrial communities alike, with applications ranging from microfluidic devices to ink-jet printing and surface coating. Though much progress has been made since the first pioneering work by Young and Laplace, many interesting, unanswered questions remain. One which has recently come to the fore because of experimental advances allowing the fabrication of surfaces with mesoscopic hydrophobic and hydrophilic regions is the behaviour of drops on chemically patterned substrates. Several authors \cite{Lenz1,Gau1,Darhuber1,Brinkmann1,Leopoldes1,Dupuis1} have shown that the wetting behaviour on these substrates can be very rich, with the drop shapes depending sensitively on parameters such as the dimensions and contact angles of the patterning. In this letter we build on this work to address the {\em dynamics} of drops moving across an array of alternating hydrophobic and hydrophilic stripes, focussing on the centre of mass motion as well as the morphological transitions induced by the imposed external flow. The drop is pushed by a constant gravity-like acceleration as opposed to \cite{Wasan1} where a thermal gradient is applied to generate the drop motion. 

We focus on three cases. First we consider the simplest case of a drop moving across a boundary between stripes of different contact angle which are much wider than the drop radius. This allows us to demonstrate the different roles of the applied force and the surface force which arise from the change in free energy as the drop crosses a stripe boundary. We then look at stripes of width of order the drop radius and choose the simulation parameters to reproduce as closely as possible the experimental values. We find pleasing agreement between simulations and experiments. We next consider a substrate  on which there is a much larger difference in contact angle between the hydrophobic and hydrophilic regions. The shape of the drop shows strong oscillations as it is pushed across this surface, and spreading across the surface perpendicular to the flow becomes important.

\section{Equations of Motion}

The equilibrium properties of the drop are described by a continuum free energy
\begin{equation} 
\Psi = \int_V (\psi_b(n)+\frac{\kappa}{2} (\partial_{\alpha}n)^2) dV
+ \int_S \psi_s(n_s) dS . \label{eq3}
\end{equation}
$\psi_b(n)$ is a bulk free energy term which we take to be \cite{Briant1}
\begin{equation}
\psi_b (n) = p_c (\nu_n+1)^2 (\nu_n^2-2\nu_n+3-2\beta\tau_w) \, ,
\end{equation}
where $\nu_n = {(n-n_c)}/{n_c}$, $\tau_w = {(T_c-T)}/{T_c}$ and $n$, $n_c$, $T$, $T_c$ and $p_c$ are the local density, critical density, local temperature, critical temperature and critical pressure of the fluid respectively. $\beta$ is a constant typically chosen to be 0.1. This choice of free energy leads to two coexisting bulk phases of density $n_c(1\pm\sqrt{\beta\tau_w})$. The second term in Eq.\ (\ref{eq3}) models the free energy associated with any interfaces in the system. $\kappa$ is related to the surface tension via $\gamma = {(4\sqrt{2\kappa p_c} (\beta\tau_w)^{3/2} n_c)}/3$ \cite{Briant1}. We take the third term as the simple Cahn expression $\psi_s (n) = -\phi \, n_s$ where $n_s$ is the value of the density at the surface \cite{Cahn}. The surface field $\phi$ can be related to the contact angle $\theta$ of the drops on the surface by 
\begin{equation}
\phi = 2\beta\tau_w\sqrt{2p_c\kappa} \,\, \mathrm{sign}(\frac{\pi}{2}-\theta)\sqrt{\cos{\frac{\alpha}{3}}(1-\cos{\frac{\alpha}{3}})} \, , \label{eq4}
\end{equation}
where $\alpha=\cos^{-1}{(\sin^2{\theta})}$ and the function sign returns the sign of its argument.

The dynamics of the drop is described by the continuity  (\ref{eq1}) and the Navier-Stokes equations (\ref{eq2})
\begin{eqnarray}
&\partial_{t}n+\partial_{\alpha}(nu_{\alpha})=0 \, , \label{eq1}\\
&\partial_{t}(nu_{\alpha})+\partial_{\beta}(nu_{\alpha}u_{\beta}) = 
- \partial_{\beta}P_{\alpha\beta}+ \nu \partial_{\beta}[n(\partial_{\beta}u_{\alpha} + \partial_{\alpha}u_{\beta} + \delta_{\alpha\beta} \partial_{\gamma} u_{\gamma}) ] 
+ na_{\alpha} \, , \label{eq2}
\end{eqnarray}
where $\mathbf{u}$, $\mathbf{P}$, $\nu$, and $\mathbf{a}$ are the local velocity, pressure tensor, kinematic viscosity, and acceleration respectively. The thermodynamic properties of the drop are input via the pressure tensor $\mathbf{P}$ which is calculated from the free energy
\begin{eqnarray} 
&P_{\alpha\beta} = (p_b(n)-\frac{\kappa}{2} (\partial_{\alpha}n)^2 -\kappa n \partial_{\gamma\gamma}n)\delta_{\alpha\beta} 
+ \kappa (\partial_{\alpha}n)(\partial_{\beta}n) \, , \nonumber \\ 
&p_b (n)= p_c (\nu_n+1)^2 (3\nu_n^2-2\nu_n+1-2\beta\tau_w) . \nonumber
\end{eqnarray}
Eqs.\ (\ref{eq1}) and (\ref{eq2}) are solved using a free energy lattice Boltzmann algorithm \cite{Briant1,Succi,Swift1}.

We consider a simulation box of size $(L_x,L_y,L_z)$. The surface at $z=0$ is patterned by (relatively) hydrophilic and hydrophobic stripes of contact angles $\theta_1$ and $\theta_2$ and widths $\delta_1$ and $\delta_2$ respectively. A drop of initial radius $R$ lies on this surface and is pushed along the $x$-direction by a constant acceleration $a_x$. No--slip boundary conditions are imposed on the velocity field at $z=0$ and $z=L_z-1$. We record the drop shape and the position of the centre of mass of the drop as a function of time. Simulation parameters which are used for all the numerical calculations are: $\kappa = 0.004$, $p_c = 1/8$, $n_c = 3.5$, $T = 0.4$, $T_c = 4/7$, and $\nu = 0.1$, while those specific to a particular simulation are given in the corresponding figure's caption.

\section{Experiments}

The experiments were performed by dispensing drops of ethylene glycol (volume $\sim 2$ mm$^{3}$, surface tension $\sigma=47.7$ mN.m$^{-1}$, dynamic viscosity $\eta=16.1$ mPa.s, and density $1.15$ g.cm$^{-3}$) on surfaces  (length = $5$ cm, width = $2$ cm) patterned by microcontact  printing. The surfaces were silicon wafers coated by a bilayer of $2$ nm of chromium and a $10$ nm layer of gold. First, a PDMS stamp was moulded on a striped master achieved by standard lithographic techniques~\cite{white1}. Then a 1mmol solution of octadecanethiol in hexane was poured onto the stamp. The stamp was applied to the surface and was followed by a quick wash with a 1 mmol solution of 1F,1F,2F,2F perfluorodecanethiol in hexane. This resulted in alternating striped layers of octadecanethiol ($0.8$ mm) and perfluorodecanethiol ($1.0$ mm) on which the contact angles of ethylene glycol are $80^{\circ}$ and $90^{\circ}$  respectively. The rms roughness of the surfaces after patterning was $\sim 2-3$ nm as determined by AFM in contact mode.

A  CCD camera (25 frames per second) linked to a computer was used to record the evolution of the drop after it had slid $3$ cm from the top of the substrate. This ensured that a stationary regime had been reached. The images were then analyzed using ImageJ software, which allowed the position of the centroid of the drop to be followed as a function of time.

\section{Results}

\begin{figure} 
\twoimages[scale=0.35,angle=0]{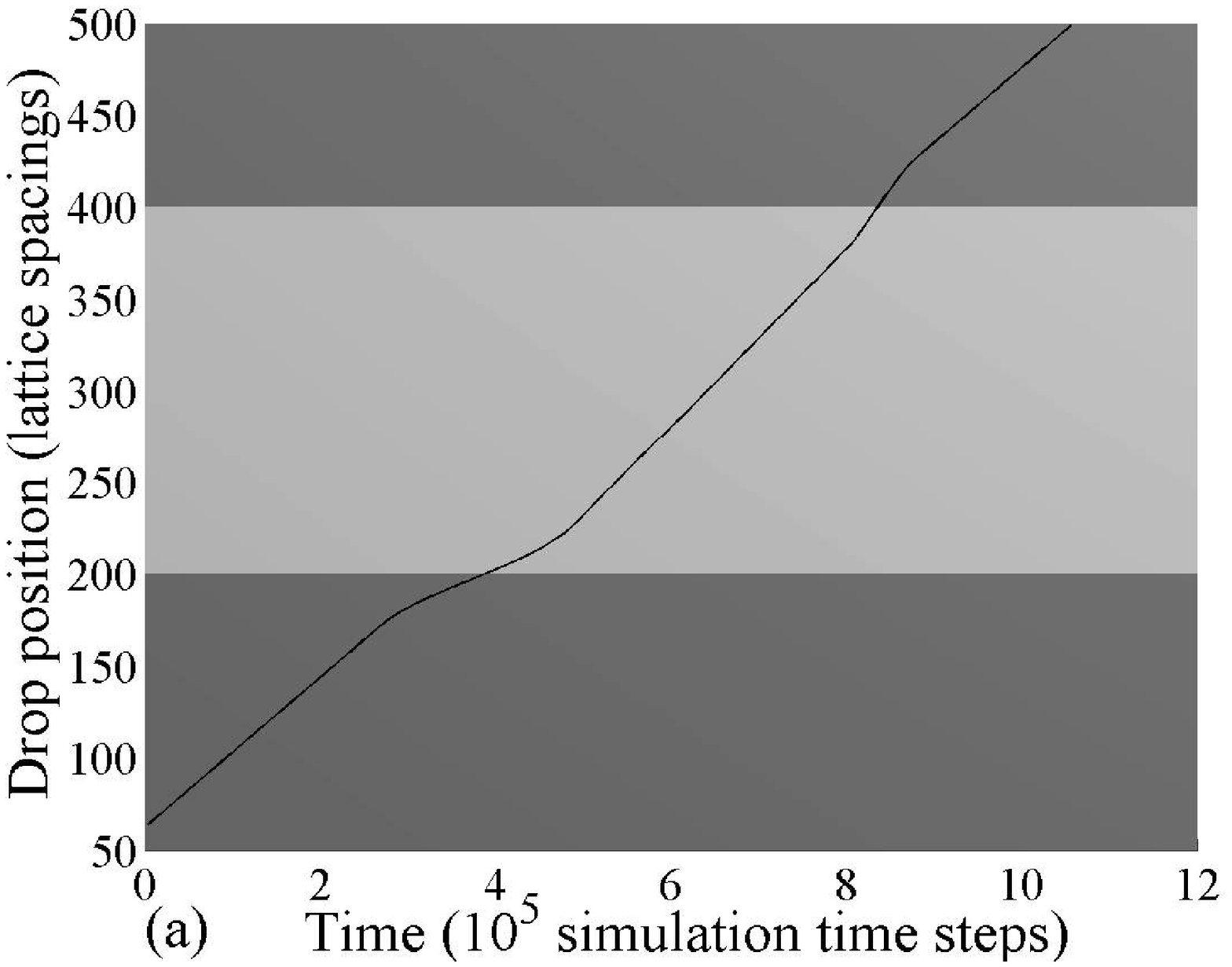}{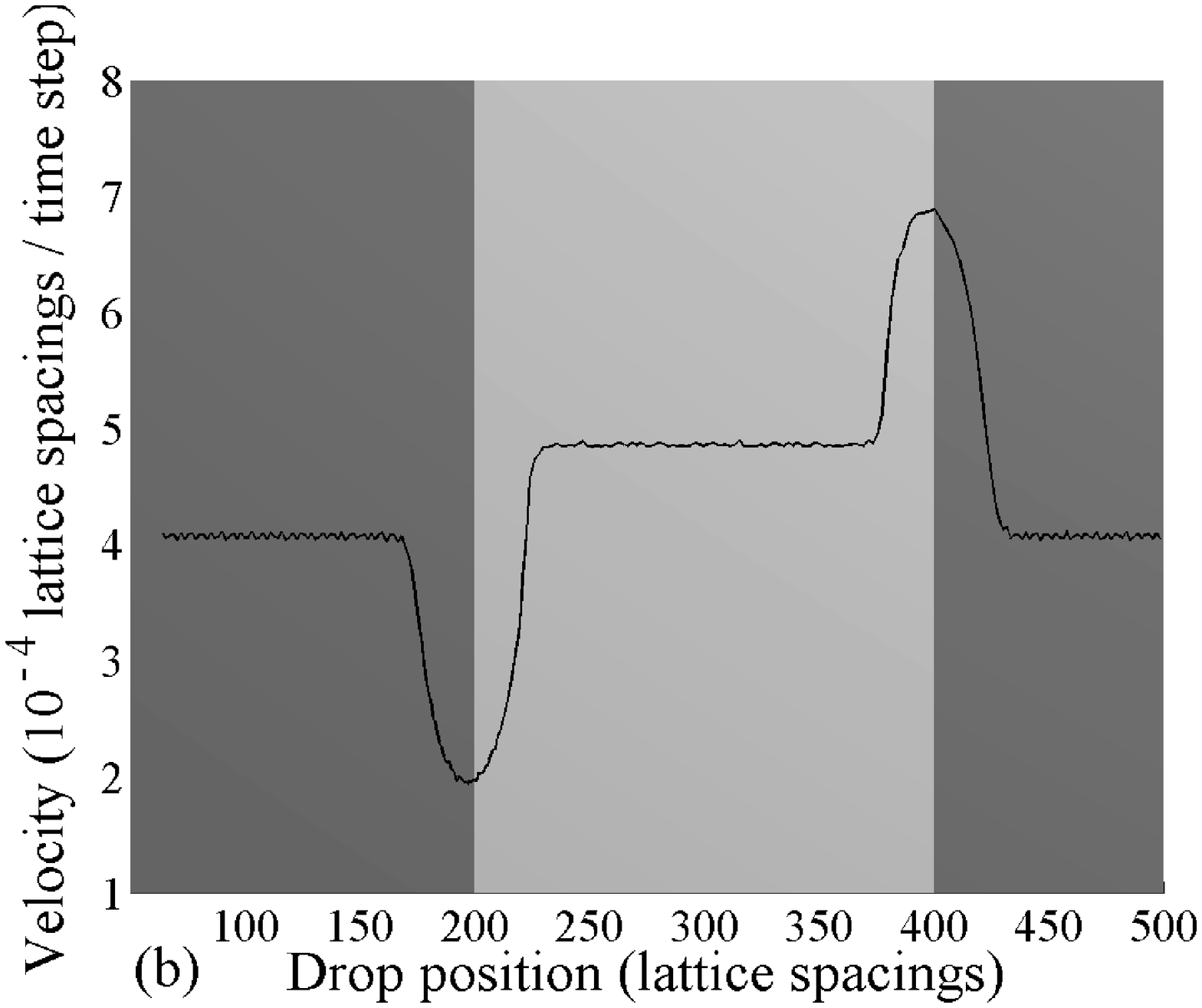}
\caption{Drop dynamics on stripes wide compared to the drop diameter: (a) drop position as a function of time. (b) drop velocity as a function of position. Simulation parameters: $(L_x,L_y,L_z) = (400,100,80)$, $\theta_1=110^{o}$ (dark grey), $\theta_2=130^{o}$ (light grey), $\delta_1/R=200/25 = 8$, $\delta_2/R=200/25 = 8$, and $a_x = 10^{-7}$.} \label{fig3:subfig}
\end{figure}

First, as a simple test case, we present numerical results for two wide stripes ($\delta_1/R = \delta_2/R = 200/25 = 8.0$) of different contact angles, $\theta_1=110^{o}$ and $\theta_2=130^{o}$. The hydrophilic and hydrophobic stripes are labelled 1 and 2 respectively. Figs.~\ref{fig3:subfig}(a) and~\ref{fig3:subfig}(b) show the drop position as a function of time, and its velocity as a function of position respectively. Away from the borders between stripes the drop attains a constant velocity as expected. It moves faster in the region of higher contact angle. This is because it is subject to a velocity profile which increases with height above the substrate; the higher the contact angle the further from the surface the drop centre of mass, and therefore the faster the drop moves.

As the drop moves from the hydrophilic to the hydrophobic stripe there is a pronounced dip in its velocity. This is due to the increase in the surface contribution to the free energy which can be estimated as
\begin{equation}
d\Psi_s = (\gamma^{SL}_2-\gamma^{SV}_2-\gamma^{SL}_1+\gamma^{SV}_1) \, y(x) dx \label{eq6}
\end{equation}
where $y(x) dx$ is the area of the base of the drop passing the chemical border as it moves through $dx$ and  $\gamma^{SL}$ and $\gamma^{SV}$ are the solid--liquid and solid--vapour surface tensions. Using Young's equation to write this in terms of contact angles gives a repulsive force
\begin{equation}
F_s = -\frac{d\Psi_s}{dx} = \gamma^{LV} (\cos{(\theta_2)}-\cos{(\theta_1)}) \, y(x)  \label{eq7}
\end{equation}
where $\gamma^{LV}$ is the liquid--vapour surface tension. The interplay between the repulsive wetting force and the increased efficiency of the applied force in moving the drop leads to the characteristic dip in fig.~\ref{fig3:subfig}(b). If the total force drops to zero we would expect the drop to remain pinned at the boundary. Note also that there is a slight increase in the width of the drop along the $y$-direction as it reaches the hydrophobic boundary, as it prefers to spread on the hydrophilic substrate rather than move across onto the hydrophobic region of the surface. However this is disfavoured by the increase in the free energy of the liquid--gas interface and is, for the small difference in surface energies considered here, a minor effect. Similarly the competition between the attractive wetting force and the decreased efficiency of the applied force account for the peak in velocity as the drop passes back into the hydrophilic region.

We next consider the case where the stripe width is smaller than the drop diameter,  ${\delta_1}/{R}=33/25=1.32$ and ${\delta_2}/{R}=27/25=1.08$, and the wettability contrast is not too large, $\theta_1=80^{o}$ and $\theta_2=90^{o}$. These parameters were chosen to match the experimental results as closely as possible. The case of a larger wettability contrast will be discussed later.

\begin{figure} 
\twoimages[scale=0.35,angle=0]{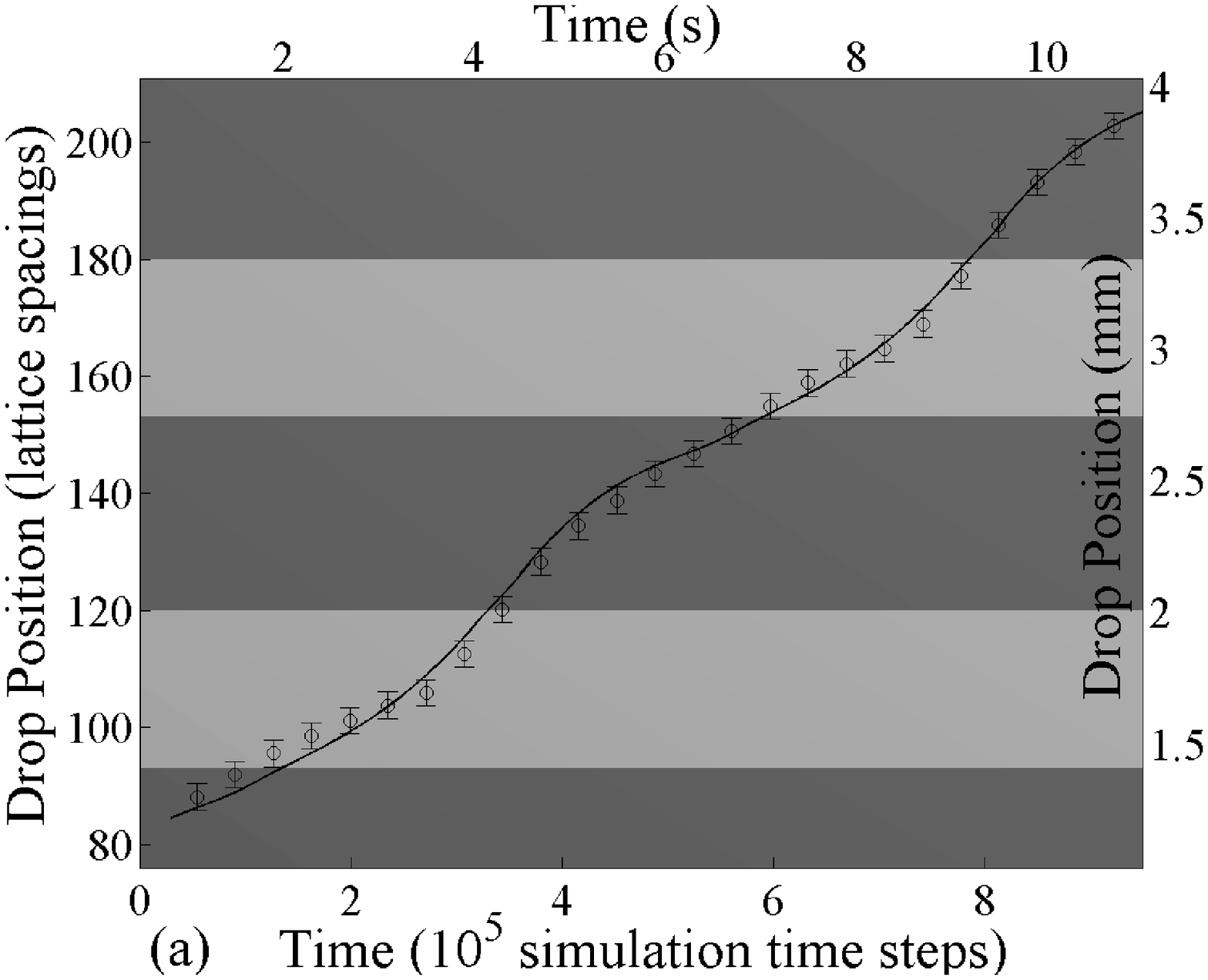}{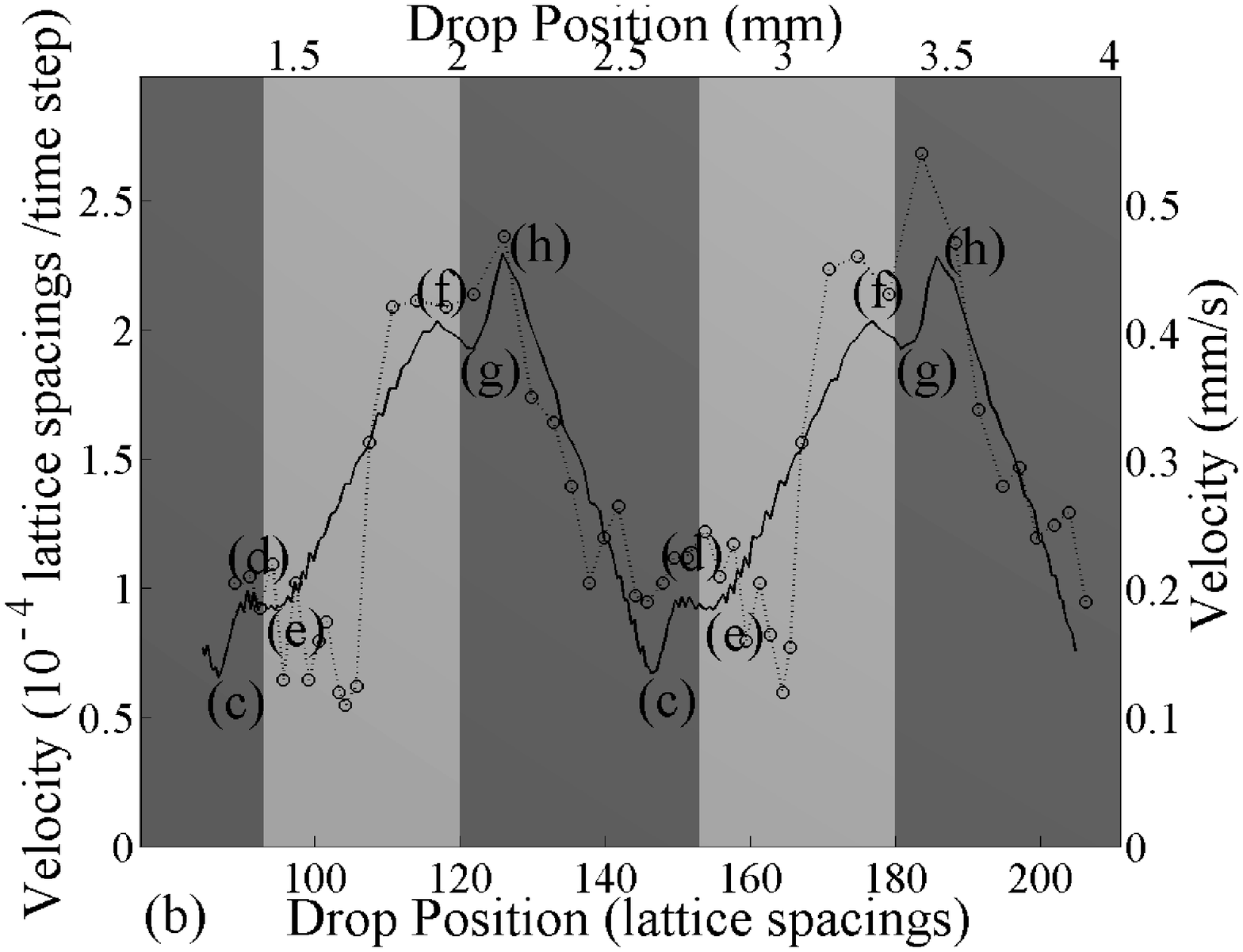}
\twoimages[scale=0.35,angle=0]{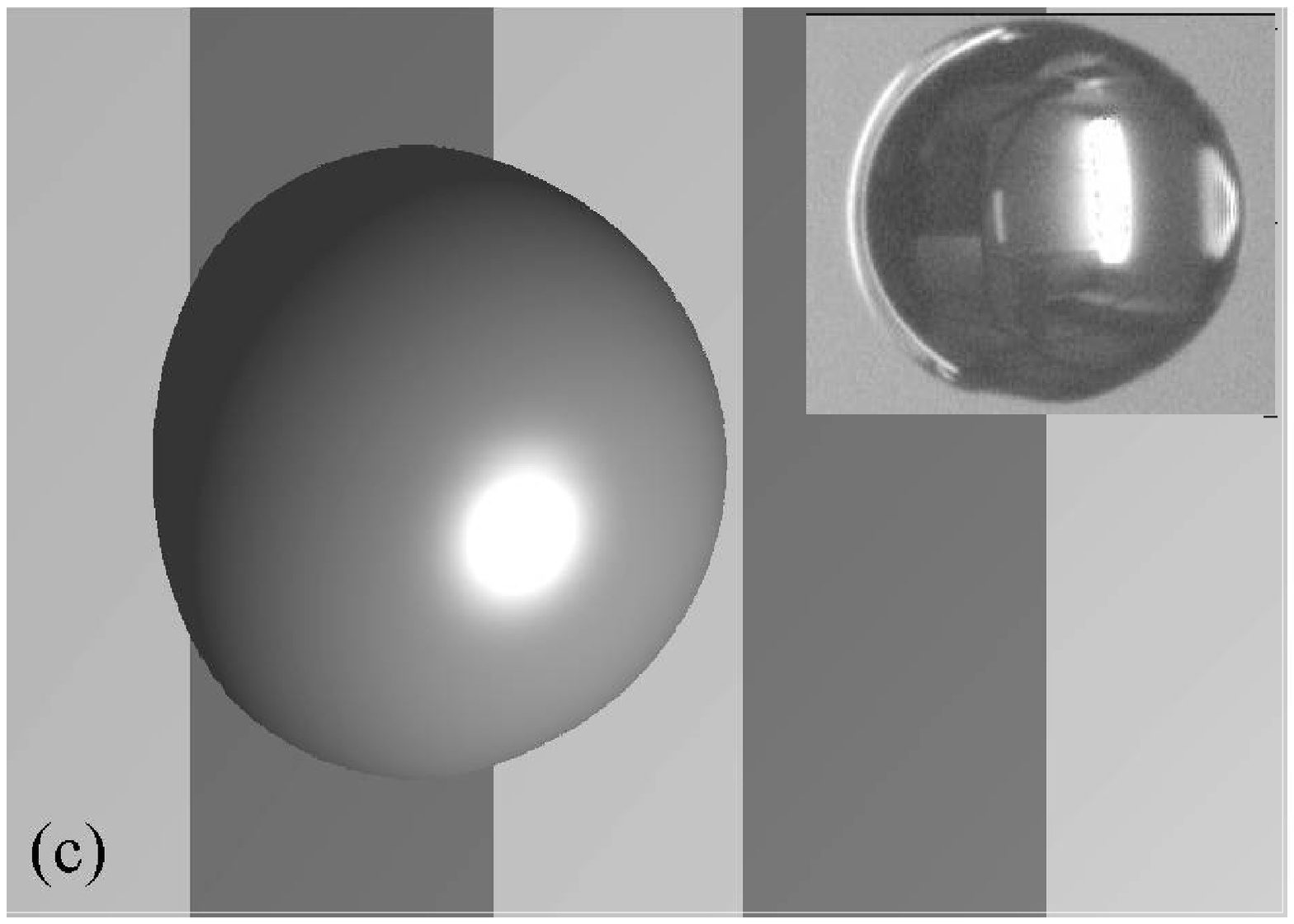}{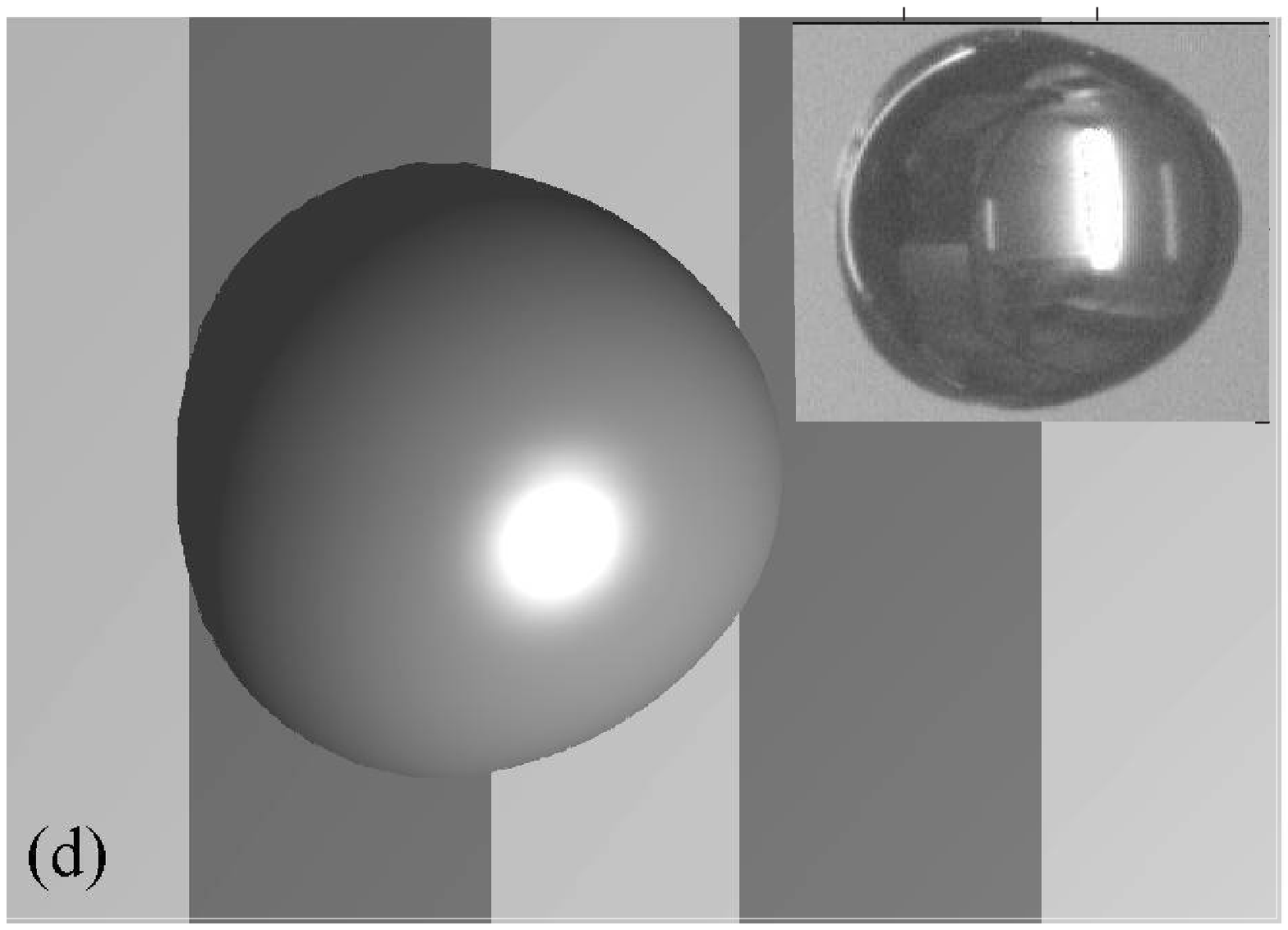}
\twoimages[scale=0.35,angle=0]{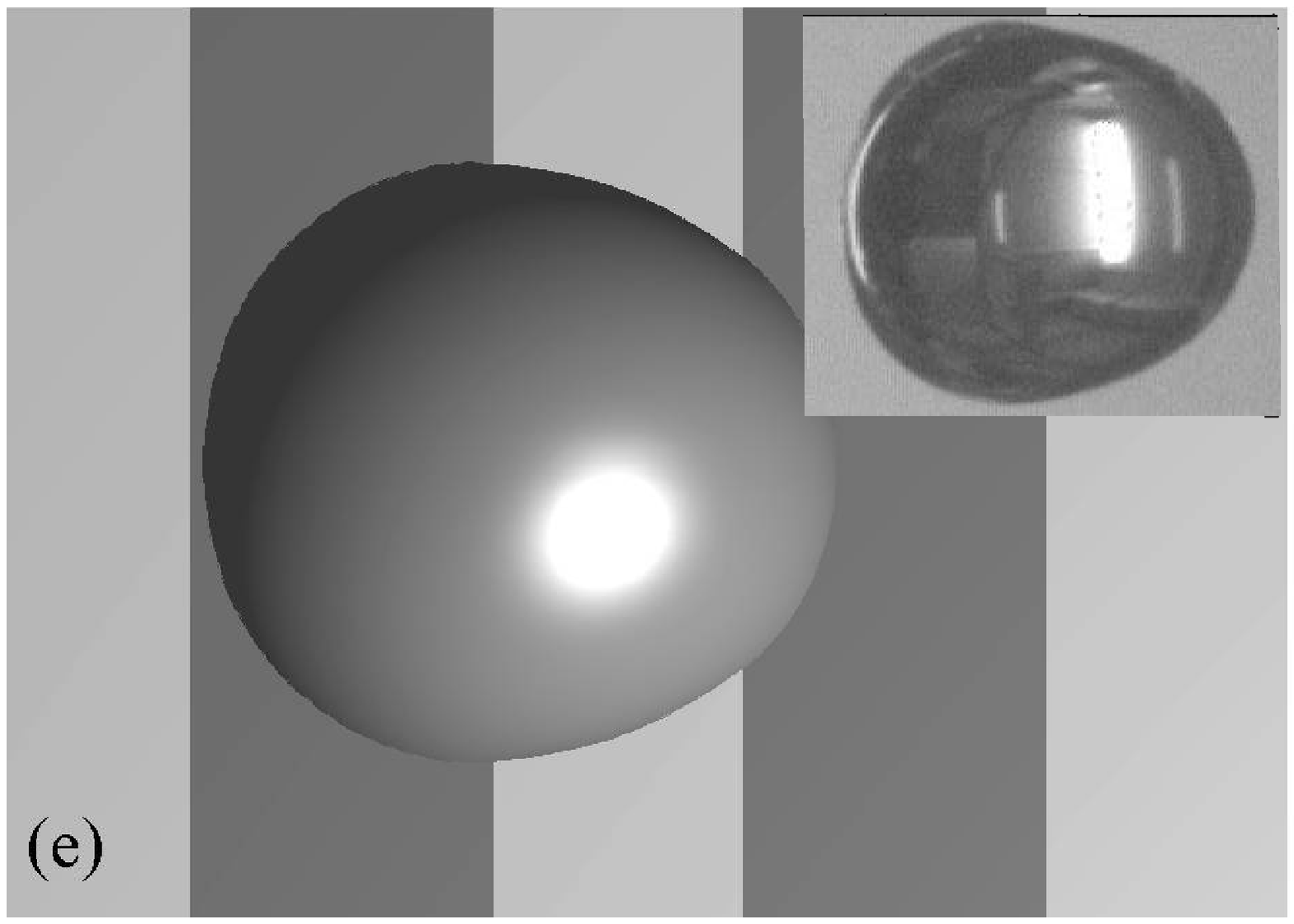}{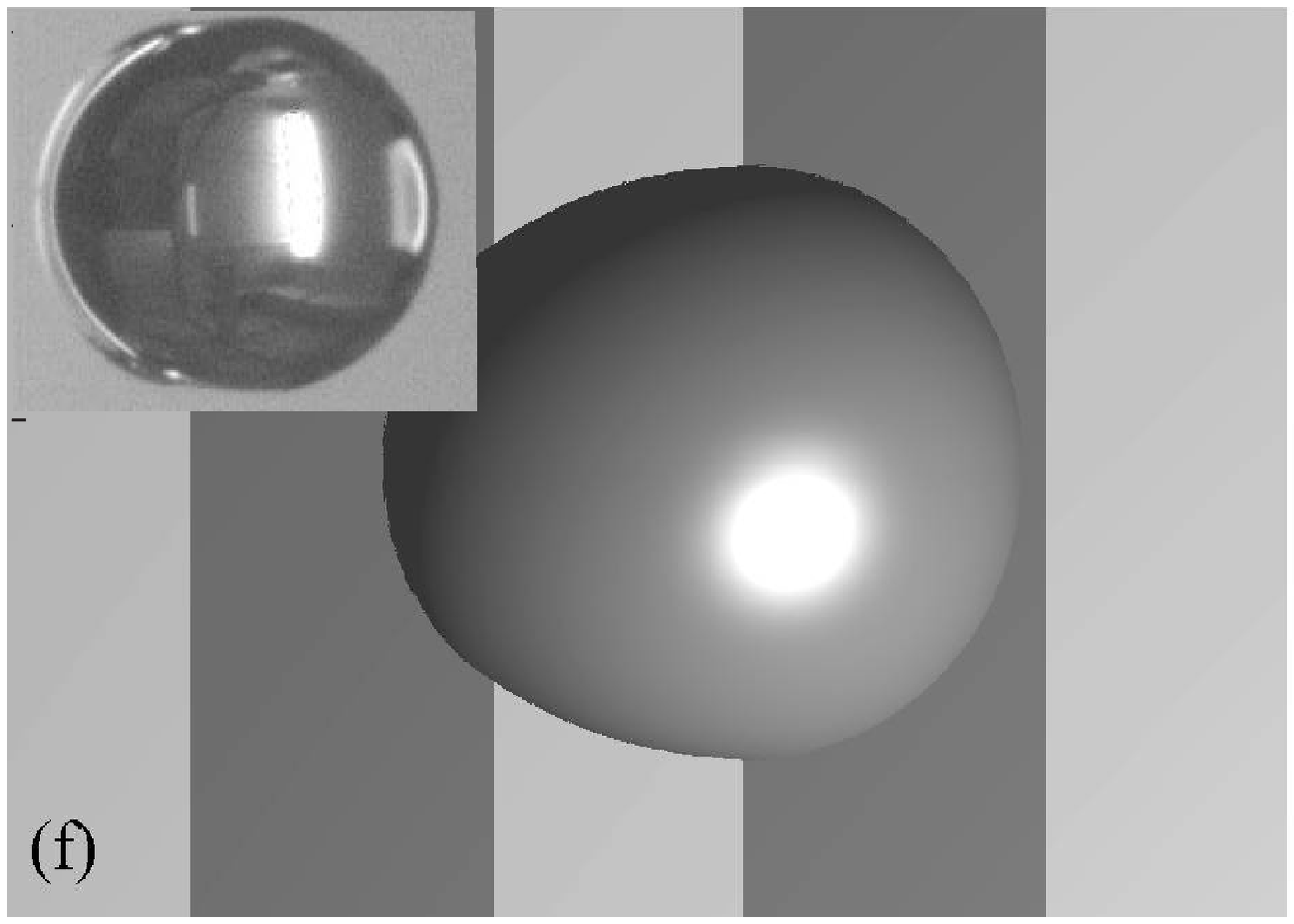}
\twoimages[scale=0.35,angle=0]{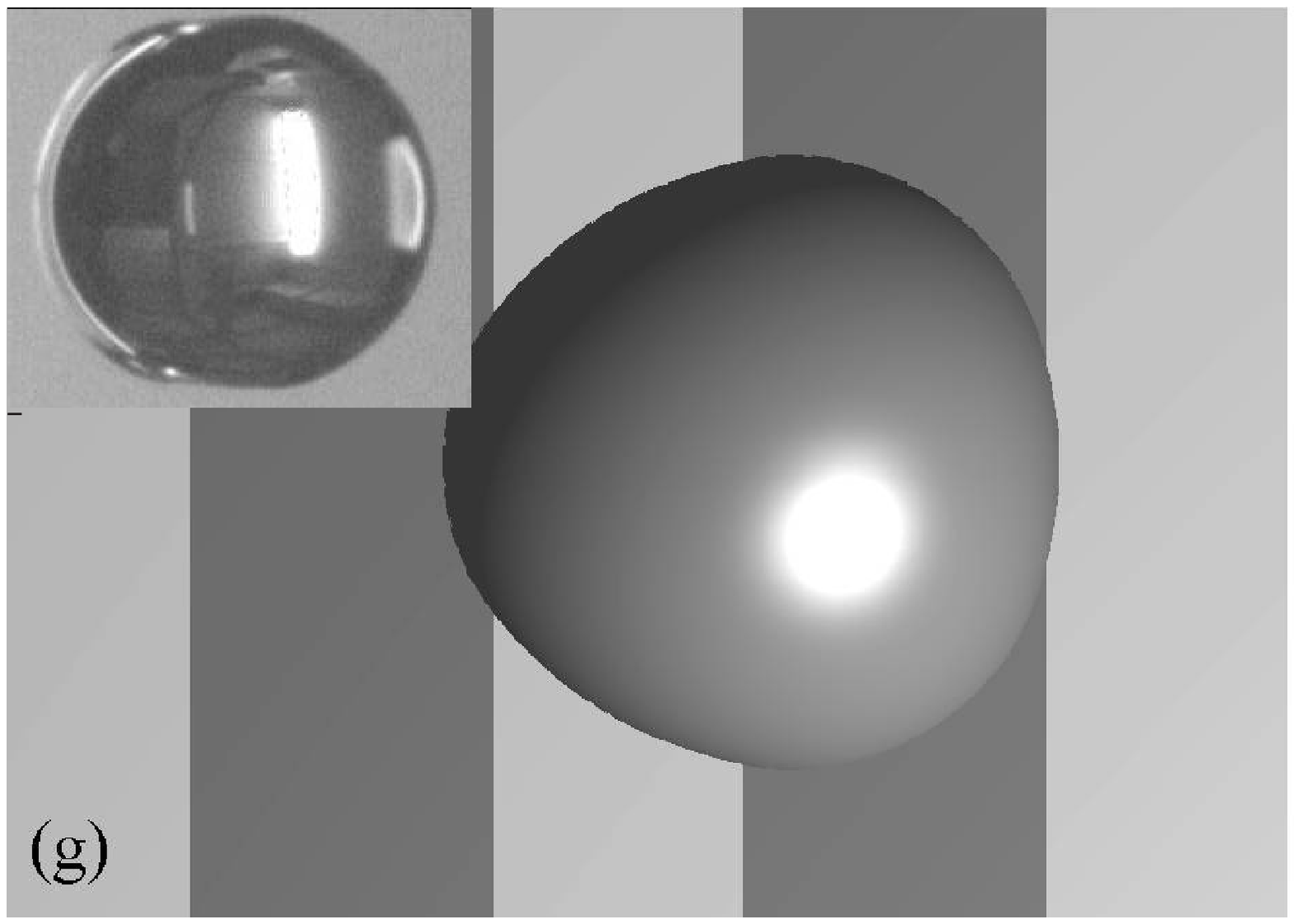}{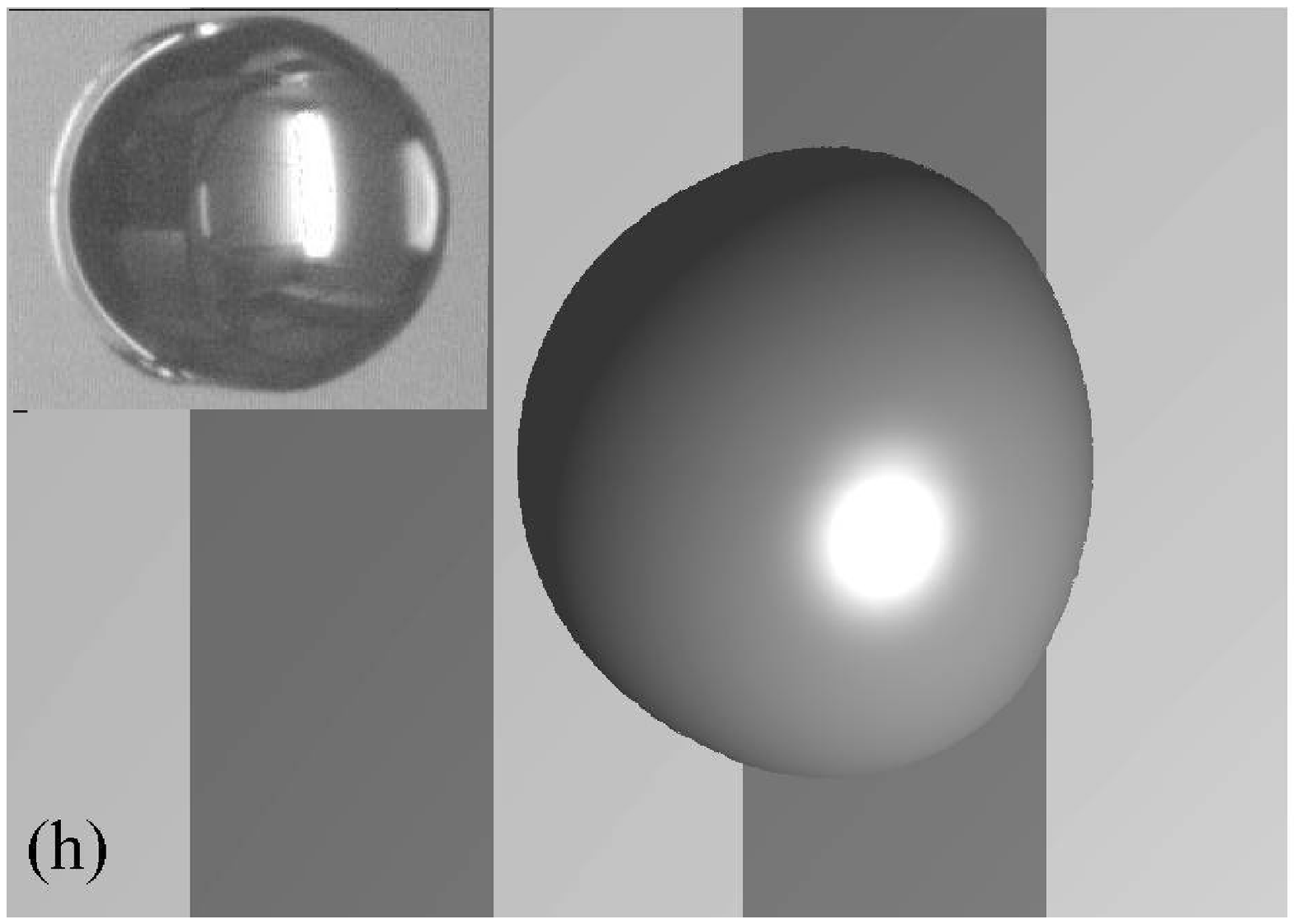}
\caption{Drop dynamics on stripes smaller than the drop diameter and with a small contact angle variation between the stripes: (a) drop position as a function of time. The continuous line shows the simulation results and the circles are the experimental results. (b) drop velocity as a function of position. (c--h) drop morphology at positions indicated in (b). The large figures are simulations and the insets are the experimental results. Simulation parameters: $(L_x,L_y,L_z) = (180,100,80)$, $\theta_1=80^{o}$ (dark grey), $\theta_2=90^{o}$ (light grey), $\delta_1/R=33/25 =1.32$, $\delta_2/R=27/25=1.08$, and $a_x = 5 \, 10^{-8}$.} 
\label{fig7:subfig}
\end{figure}

We first describe the numerical results. In fig.~\ref{fig7:subfig}(a) the drop position is plotted as a function of time and fig.~\ref{fig7:subfig}(b) shows the drop velocity as a function of its position on the substrate. We also show, in figs.~\ref{fig7:subfig}(c--h), snapshots of the drop shape at designated points along its path. The first difference, when compared to the wider stripes, is that the drop never reaches a constant velocity. However, as expected, the periodicity of the substrate is reflected in the variaton of the drop velocity with time. The broad features of the curves match those in fig.~\ref{fig3:subfig}(b) with the flat sections removed showing that the dynamics are again controlled by an interplay between the applied and wetting forces. The additional structure at the peaks and troughs of the velocity versus position graph occurs because the drop is traversing two stripe boundaries simultaneously.

We now compare to the experimental results.  The simulation parameters were chosen so that the values of the liquid viscosity, the surface tension, the drop radius and the stripe widths in the simulations matched the experimental values. We also chose the (static) contact angles to take values of $80^{o}$ and $90^{o}$.

The drop displacement as a function of time measured in the experiments is shown by the data points in fig.~\ref{fig7:subfig}(a). Photographs of the drop as it moves across the substrate are shown as insets in figs.~\ref{fig7:subfig}(c--h). There is a pleasing agreement between simulations and experiments.

However we caution that, as with all mesoscale approaches, it is not possible to match all the physical parameters and retain a stable and numerically feasible calculation. The interface is too wide in the simulations (relative to the drop radius) and the liquid--gas density difference is too small. The contact line velocity $u \propto \xi/(\Delta n)^2$, where $\xi$ is the interface width and $\Delta n$ is the density difference \cite{Briant1}. The drop therefore moves too quickly in the simulations. This problem is accounted for by rescaling the time axis. Another difference between the simulations and experiments is that experimental drops are not confined in a channel as in the simulations. However, we have chosen $L_z/R$ sufficiently large that the drop feels a parabolic velocity profile akin to the experimental profile.

\begin{figure} 
\twoimages[scale=0.35,angle=0]{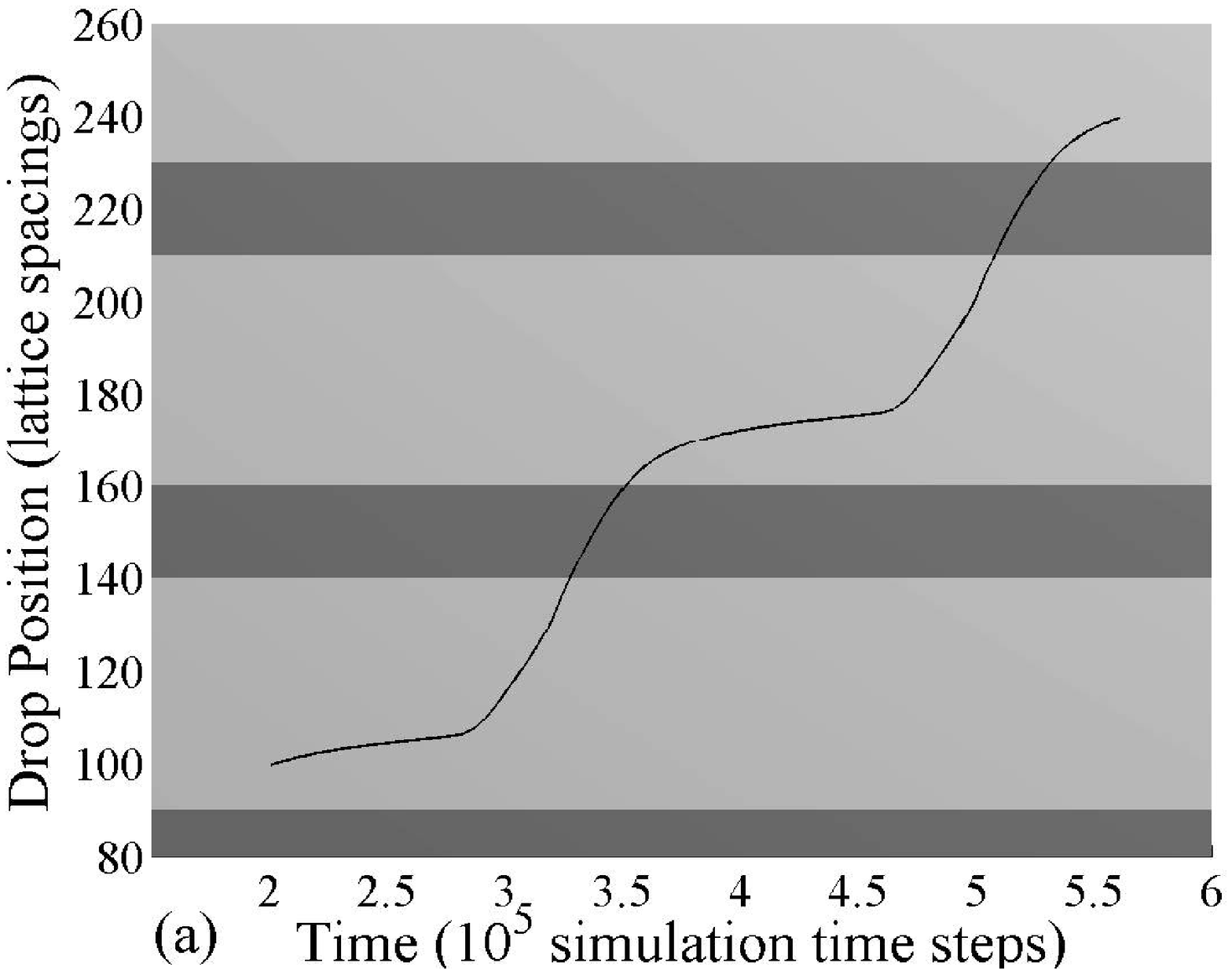}{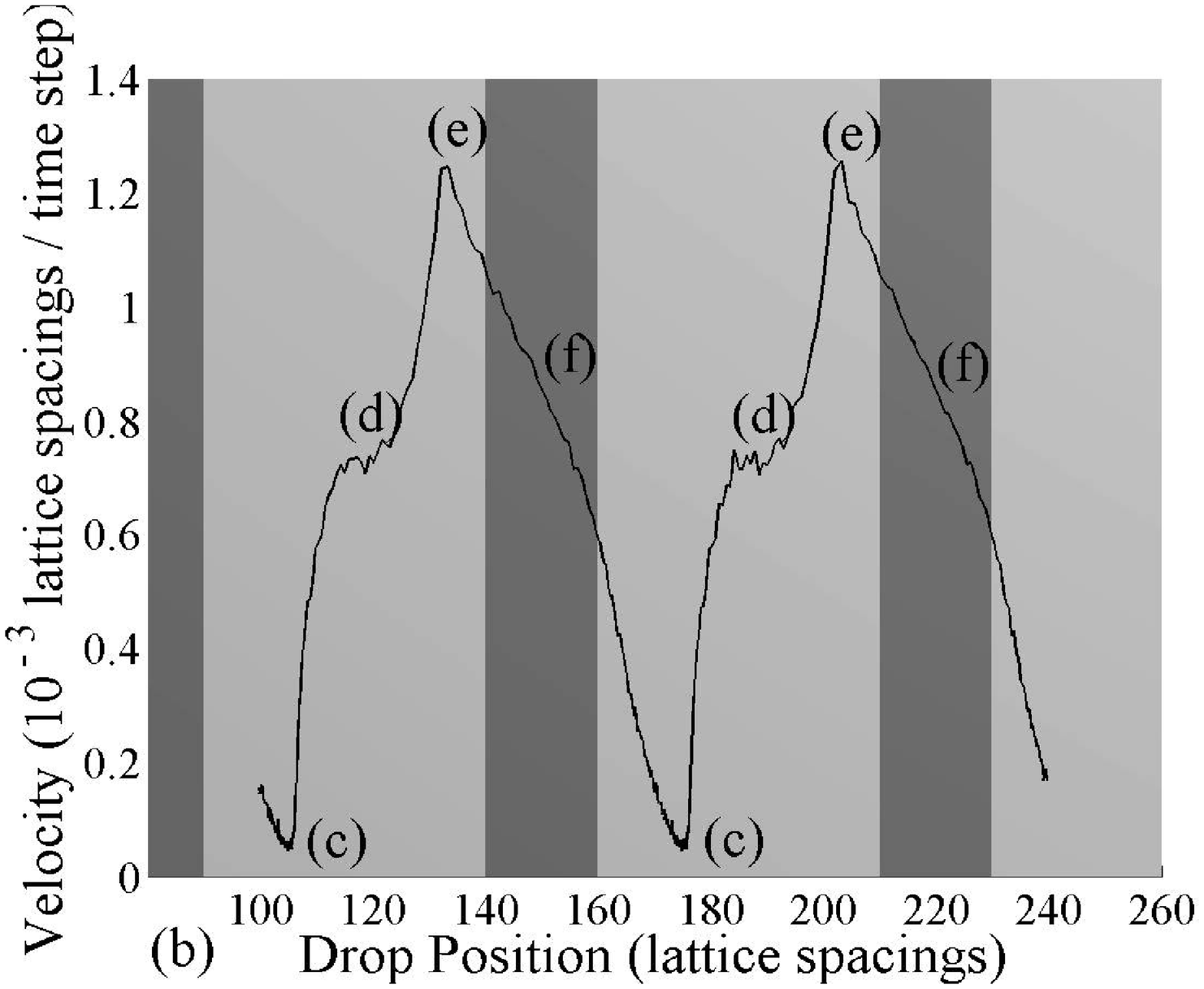}
\twoimages[scale=0.41,angle=0]{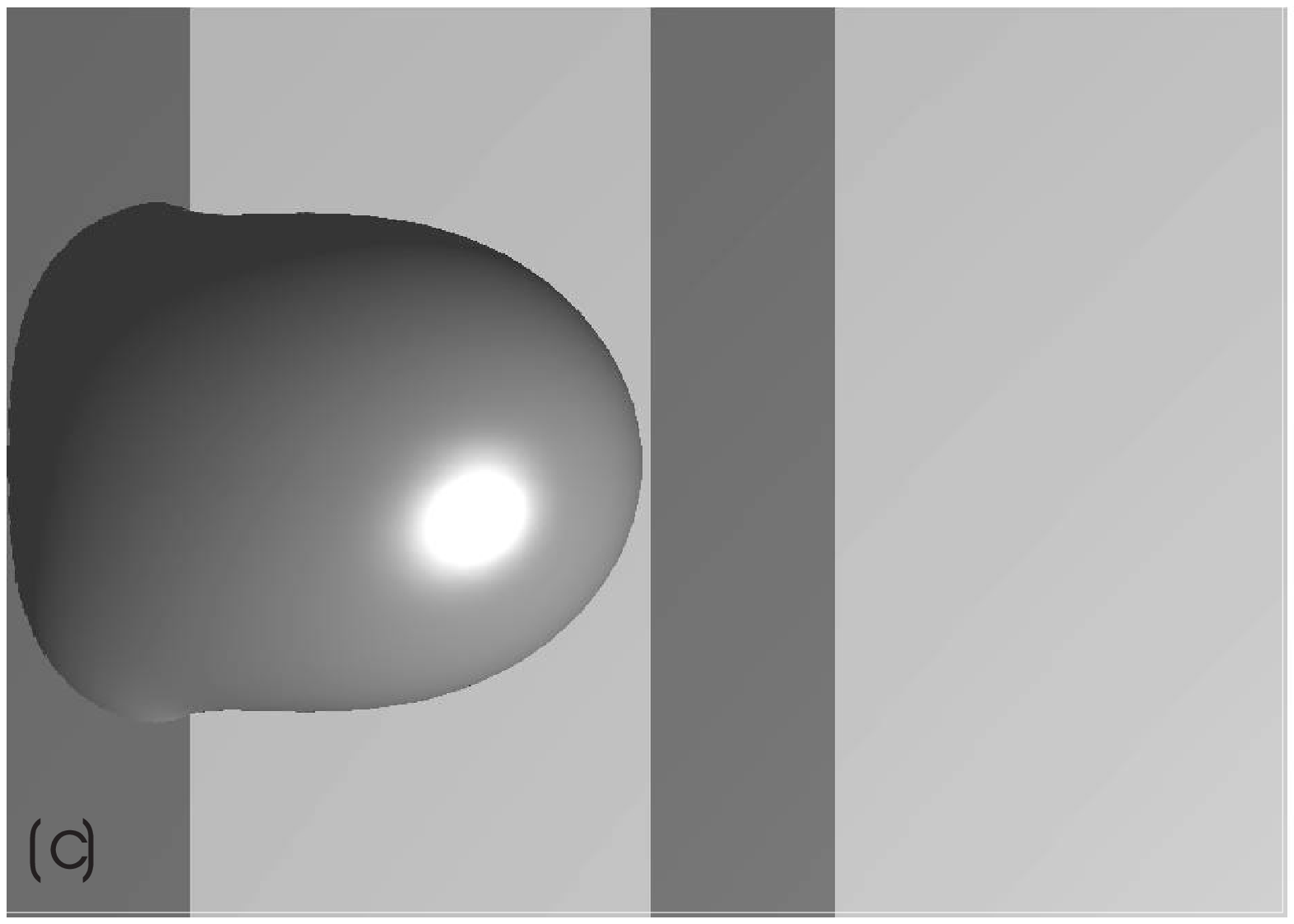}{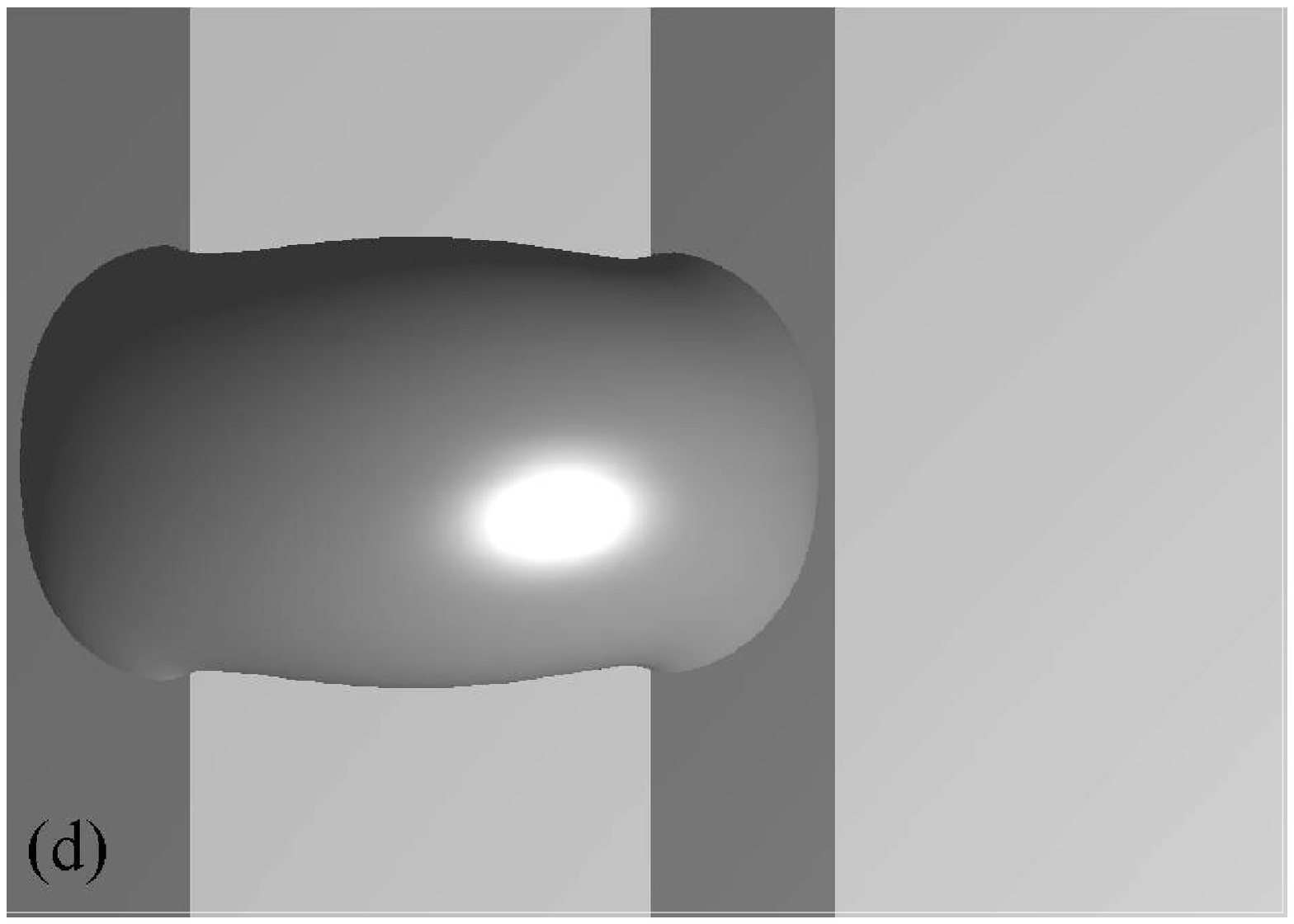}
\twoimages[scale=0.41,angle=0]{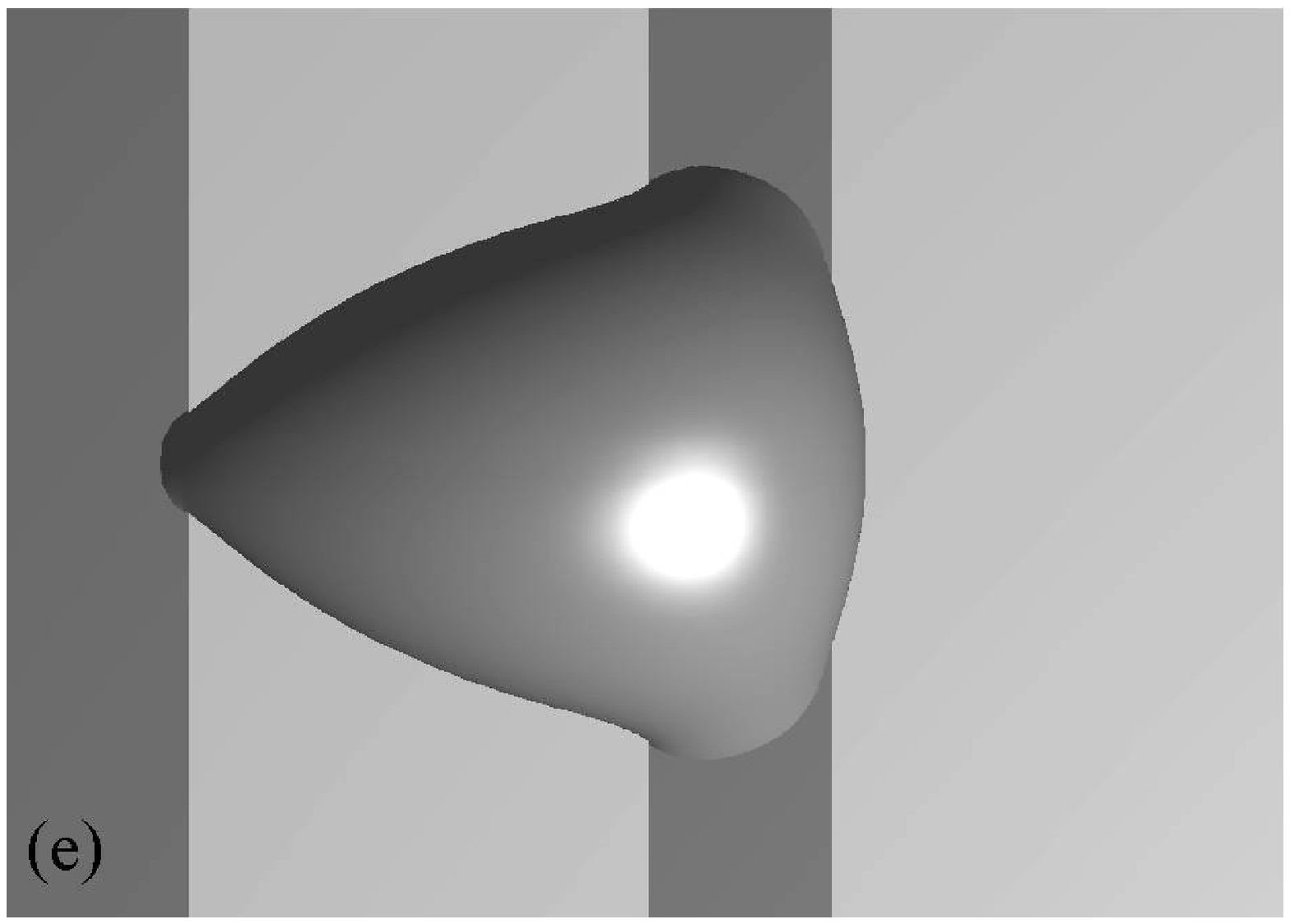}{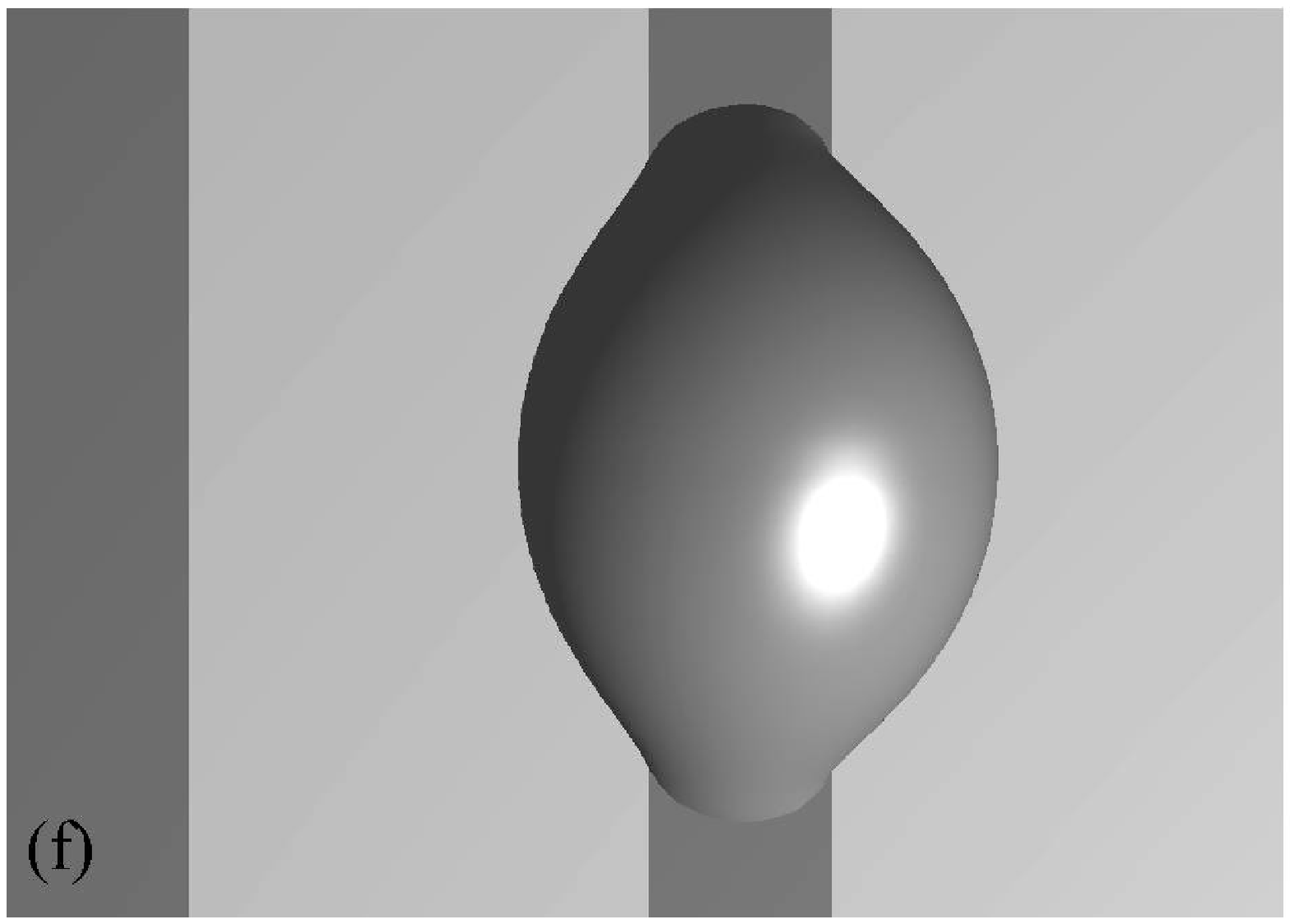}
\caption{Drop dynamics where there is a large contact angle variation between the stripes: (a) drop position as a function of time. (b) drop velocity as a function of position. (c--f) drop morphology at positions indicated in (b). Simulation parameters: $(L_x,L_y,L_z) = (140,100,80)$, $\theta_1=60^{o}$ (dark gray), $\theta_2=110^{o}$ (light gray), $\delta_1/R=20/25 = 0.8$, $\delta_2/R=50/25=2.0$, and $a_x = 2.5 \, 10^{-7}$.} 
\label{fig8:subfig}
\end{figure}

However, given these caveats, the agreement between simulation and experiment gives us confidence to explore the case of a larger contact angle difference between stripes which is more difficult to achieve experimentally. We take ${\delta_1}/{R}=0.8$, ${\delta_2}/{R}=2.0$, $\theta_1=60^{o}$ and $\theta_2=110^{o}$. We plot the drop position as a function of time and the drop velocity as a function of position in figs.~\ref{fig8:subfig}(a) and~\ref{fig8:subfig}(b) respectively. Several authors \cite{Brinkmann1,Leopoldes1,Dupuis1} have shown that on such a substrate the drop can take two final shapes. One is the `butterfly' configuration where the drop spans the space between two hydrophilic stripes (eg fig.~\ref{fig8:subfig}(d)) and the second is the `diamond' shape where the drop just lies on a single hydrophilic stripe (eg fig.~\ref{fig8:subfig}(f)). Strong deviations from spherical can now be favourable because the extra liquid--gas interface free energy is offset by the considerable advantage of lying on a hydrophilic stripe. 

When a constant body force is applied, the drop moves across the
surface and changes from a diamond to a butterfly shape and back again
(fig.~\ref{fig8:subfig}(c--f)). Let us assume the drop is initially in
a diamond configuration (fig.~\ref{fig8:subfig}(f)). Due to the
Poiseuille flow field, the drop is pushed forward onto the hydrophobic
region. Its velocity decreases because of the dewetting force at the
hydrophobic stripe. If the external body force is small, the drop
velocity will fall to zero, the drop will be pinned, and the
steady-state drop shape will look similar to
fig.~\ref{fig8:subfig}(c). For the parameters we consider here,
however, the drop is just able to channel to the next hydrophilic
stripe. The effective capillary force at the hydrophobic-hydrophilic
border then starts to take charge and the drop accelerates and wets
the next hydrophilic stripe. The drop now has a butterfly
shape (fig.~\ref{fig8:subfig}(d)) and is at the peak of an energy
barrier between two diamond shapes on successive hydrophilic
stripes. It then becomes more advantageous for the drop to spread along
the new hydrophilic stripe in the $y$-direction than to continue to move
along the substrate. Hence the diamond config\
uration is re-formed and the oscillations repeat.

\section{Summary}

We have presented numerical and experimental results following the dynamics of micron--scale drops moving across a chemically patterned surface. The drop behaviour is determined by the balance between the driving force and the variation in the surface force as it moves between regions of different wettability. As the wettability contrast increases there are large deviations of the drop shape from spherical as it moves across the surface.

We have shown that there is close agreement between results of the lattice Boltzmann simulations and experiments. Therefore we hope that the numerical approach will provide a useful tool for designing new ways of controlling the motion of liquid drops on patterned surfaces for applications in areas such as microfluidics and ink-jet printing.

\acknowledgements

We thank D.G. Bucknall for useful discussions. AD and JL acknowledge the support of the EC IMAGE-IN project GR1D-CT-2002-00663 and HK support from a Clarendon Bursary and the Tanoto foundation.

\end{document}